\documentclass[amssymb, superscriptaddress, twocolumn, aps, prl,longbibliography,nobibnotes]{revtex4-2}

\setcitestyle{super}
\usepackage{graphicx}
\usepackage{amsmath}
\usepackage{upgreek}
\usepackage{color}
\usepackage{xcolor}
\definecolor{navy}{rgb}{0,0,0.80}
\usepackage{setspace}
\usepackage{hyperref}
\hypersetup{colorlinks,breaklinks,
            urlcolor=[rgb]{0,0,0.80},
            linkcolor=[rgb]{0,0,0.80},
            citecolor=[rgb]{0,0,0.80}}
\usepackage{nccmath}
\usepackage{amssymb}
\usepackage{bbding}

\newcommand{\Cal}{Department of Chemistry, University of California, Berkeley, CA 94720, USA. }
\newcommand{\LBNL}{Materials Sciences Division, Lawrence Berkeley National Laboratory, Berkeley, CA 94720, USA. }

\newcommand{\OSU}{Department of Chemistry and Biochemistry, The Ohio State University, Columbus, OH 43210, USA. }
\newcommand{\caltech}{Division of Chemistry and Chemical Engineering, California Institute of Technology, Pasadena, CA 91125, USA. }

\newcommand{\argonne}{Advanced Photon Source, Argonne National Laboratory, 9700 South Cass Avenue, Argonne, IL 60439, USA. }
\newcommand{\argonneMSD}{Materials Science Division, Argonne National Laboratory, 9700 South Cass Avenue, Argonne, IL 60439, USA.}

\newcommand{\papertitle}{Hidden correlations in stochastic photoinduced dynamics of a solid-state electrolyte}

\begin{document}
\thispagestyle{empty}

\onecolumngrid
\begin{center}
\textbf{\large \papertitle}
\vspace{0.25cm}
\end{center}

\begin{center}
    
Jackson~McClellan,$^{1,\,2,\,3,\,{\color{navy}\ast}}$ Alfred~Zong,$^{1,\,3,\,{\color{navy}\ast,\,\text{\Envelope}}}$ Kim~H.~Pham,$^{4}$ Hanzhe~Liu,$^{4}$ Zachery~W.~B.~Iton,$^{4}$ Burak~Guzelturk,$^{5}$ Donald~A.~Walko,$^{5}$ 
Haidan~Wen,$^{5,\,6}$ Scott~K.~Cushing,$^{4,\,{\color{navy}\text{\Envelope}}}$ Michael~W.~Zuerch,$^{1,\,3,\,{\color{navy}\text{\Envelope}}}$

\vspace{0.2cm}
(Dated: \today)
\end{center}

\begin{small}
\begin{singlespace}
{\it
\noindent$^1$\Cal

\noindent$^2$\OSU

\noindent$^3$\LBNL

\noindent$^4$\caltech

\noindent$^5$\argonne

\noindent$^6$\argonneMSD
}

\noindent$^{\color{navy}\ast\,}$These authors contributed equally to this work: Jackson~McClellan and Alfred~Zong.

\noindent$^{\color{navy}\text{\Envelope}\,}$Correspondence to: A.Z. (\href{mailto:alfredz@berkeley.edu}{alfredz@berkeley.edu}), S.K.C. (\href{mailto:scushing@caltech.edu}{scushing@caltech.edu}), and M.W.Z. (\href{mailto:mwz@berkeley.edu}{mwz@berkeley.edu}).
\end{singlespace}
\end{small}
\vspace{0.4cm}

\begin{quote}
\begin{small}
\textbf{Abstract}: Photoexcitation by ultrashort laser pulses plays a crucial role in controlling reaction pathways, creating nonequilibrium material properties, and offering a microscopic view of complex dynamics at the molecular level. The photo response following a laser pulse is, in general, non-identical between multiple exposures due to spatiotemporal fluctuations in a material or the stochastic nature of dynamical pathways. However, most ultrafast experiments using a stroboscopic pump-probe scheme struggle to distinguish intrinsic sample fluctuations from extrinsic apparatus noise, often missing seemingly random deviations from the averaged shot-to-shot response. Leveraging the stability and high photon-flux of time-resolved X-ray micro-diffraction at a synchrotron, we developed a method to quantitatively characterize the shot-to-shot variation of the photoinduced dynamics in a solid-state electrolyte. By analyzing temporal evolutions of the lattice parameter of a single grain in a powder ensemble, we found that the sample responses after different shots contain random fluctuations that are, however, not independent. Instead, there is a correlation between the nonequilibrium lattice trajectories following adjacent laser shots with a characteristic ``correlation length'' of approximately 1,500 shots, which represents an energy barrier of 0.38~eV for switching the photoinduced pathway, a value interestingly commensurate with the activation energy of lithium ion diffusion. Not only does our nonequilibrium noise correlation spectroscopy provide a new strategy for studying fluctuations that are central to phase transitions in both condensed matter and molecular systems, it also paves the way for discovering hidden correlations and novel metastable states buried in oft-presumed random, uncorrelated fluctuating dynamics.
\end{small}
\end{quote}

\vspace{0.4cm}

\twocolumngrid

\begin{figure*}[t!]
    \includegraphics[width=0.95\textwidth]{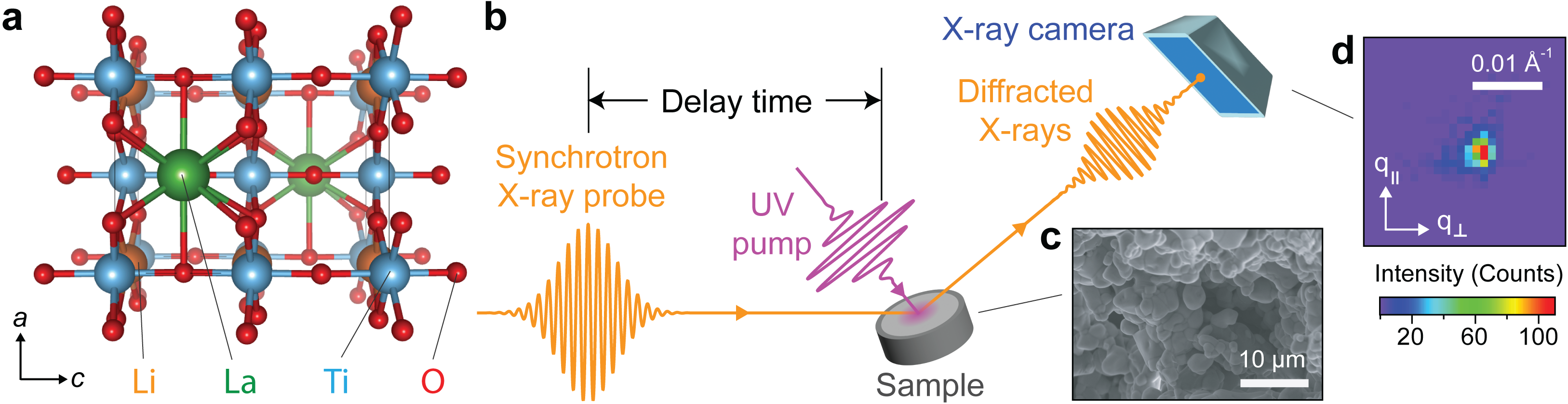}
    \caption{\textbf{Overview of the powder sample and the pump-probe setup.}  \textbf{a},~Crystal structure of the tetragonal LLTO in equilibrium. Graphics rendered by VESTA \cite{Momma2011}.  \textbf{b},~Schematic of the experimental setup for synchrotron-based time-resolved hard X-ray micro-diffraction (see Methods). \textbf{c},~Scanning electron micrograph of the  powder LLTO sample pressed into a pellet, which was used in the time-resolved X-ray experiment. \textbf{d},~Raw diffraction image of the (004) Bragg peak of a particular crystalline grain recorded before pump pulse arrival. $\mathbf{q}_\perp$ is along the $c$-axis of the grain while $\mathbf{q}_\parallel$ lies in the $a$-$b$ plane of the grain.}
    \label{fig:1}
\end{figure*}

Even though rapid advances in laser technology have enabled the detection of ultrafast dynamics in biological systems \cite{Moffat1998}, quantum materials \cite{DelaTorre2021,Zong2023}, and chemical reactions \cite{Lepine2014} down to the attosecond regime, probing stochastic processes that are intrinsic in these systems remains a conundrum. In cases where random fluctuations and nondeterministic quasiparticle motions dominate the dynamics of interest, such as near the critical point of a phase transition or in a disordered medium riddled with topological defects, a conventional stroboscopic pump-probe scheme is ineffective as it averages out shot-to-shot variations in the pump-induced response. Even if the probe signal achieves a sufficiently high signal-to-noise ratio in a non-stroboscopic single-shot pump-probe measurement \cite{Mo2016,Liang2018,Helk2019}, typically only the data point at one pump-probe delay is collected after a single pump shot, rendering it impossible to reconstruct the stochastic dynamics over time. Alternatively, a pump pulse may be followed by a train of probe pulselets via echelon-based optics \cite{Wakeham2000,Shin2014,Noe2016} to characterize the entire pump-induced evolution. However, the probe signal spread over individual pulselets can be too weak to yield useful information about the dynamics after only one pump shot, and extensive averaging over multiple pump shots are often necessary. Fundamentally, there is an upper limit on how strong the probe pulse (or pulselets) can be so that it does not significantly alter the dynamics under investigation. This limit hence imposes a trade-off between the signal-to-noise ratio and how many delay times can be measured, preventing access to the full, stochastic dynamics that differ from one pump shot to another.

Here, we tackle this challenge in studying stochastic ultrafast dynamics via a new statistical approach that is capable of uncovering hidden correlations between individual pump-induced events. A key to this method relies on the low noise level associated with the data collection process that can be plagued by instrumental instability or low photon counts in a table-top ultrafast laser-based measurements. In our experiment, we overcome these problems by utilizing the stability and high photon flux of a synchrotron-based time-resolved X-ray setup such that the experimentally observed signal variation is dominated by the intrinsic sample fluctuations in the photoinduced response. Unlike previous approaches that necessitate a single-shot setup \cite{Klose2023,Mangu2024}, our method is deployed to a typical stroboscopic pump-probe scheme and only necessitates repeated data acquisitions of the same pump-probe delay with a stable light source and high measurement statistics. We term this method nonequilibrium noise correlation spectroscopy, which can have broad applicability to a wide range of time-resolved experiments. We expect this approach to yield a deeper understanding of the key role of disordered fluctuations not just in lattice but also in charge, spin, and orbital degrees of freedom that are inherent in the nonequilibrium photoinduced trajectories \cite{Wall2018,Zong2021b,Perez2022,Disa2023}.

To demonstrate this capability of unraveling dynamical correlations in the transient stochastic response, we chose to study Li$_{0.5}$La$_{0.5}$TiO$_3$ (LLTO), a highly conductive solid-state electrolyte candidate for lithium-ion batteries \cite{Stramare2003, Belous2004, Nakayama2005, Qian2012}. LLTO has a perovskite crystal structure (Fig.~\ref{fig:1}a) with alternating vacancy-rich and poor layers \cite{Okumura2011}, where the low diffusion barrier leads to high lithium ion conductivity \cite{Zhang2020}. The mobility of lithium ions via LLTO is highly dependent on the crystal structure \cite{Woodahl2023}, which changes substantially upon lithium ion insertion and extraction as observed in X-ray diffraction \cite{Zhang2020}. Conversely, if the crystal structure changes upon photoexcitation due to nonthermal phonon population \cite{RenedeCotret2019,Cheng2024} and transient heat deposition \cite{Sokolowski1995,Shugaev2016}, one also expects varying dynamics from shot to shot due to the stochastic location of the lithium ions in the crystal, which are free to migrate either due to thermal activation \cite{Yashima2005} or photoexcitation \cite{Kim2023}. Therefore, the highly mobile lithium ions in LLTO offer the ideal platform to investigate stochastic fluctuations in the lattice dynamics following photoexcitation, which can in turn yield insight into the strong coupling between lithium ion diffusion and the crystalline lattice \cite{Woodahl2023,Wakamura1998} for designing next-generation solid-state electrolyte.

To this end, we monitored the evolution of lattice parameters of LLTO following above-bandgap photoexcitation using synchrotron-based X-ray diffraction (see Methods for details). An above-bandgap excitation in LLTO is expected to cause heat-induced initial lattice expansion and subsequent relaxation that modulate lithium site occupations because it is known that the activation energy of ionic transport is closely tied to lattice distortions \cite{Wakamura1998}. Importantly, we uncovered a hidden correlation between lattice motions following neighboring pump shots with a characteristic ``correlation length'' of approximately 1,500 laser shots, which correspond to a $\sim$0.38~eV energy barrier. The energy barrier for lithium ion conduction in LLTO is similar in magnitude, so the stochastic lattice trajectories may give insight into the photoswitched ion diffusion mechanism. The novel framework of nonequilibrium noise correlation spectroscopy introduced in this work provides new understanding in the potential role of photo-induced structural change in the lithium-ion conduction process, opening the avenue for harnessing tailored laser pulses for manipulating ionic conduction in solids.

\begin{figure*}[t!]
    \includegraphics[width=0.94\textwidth]{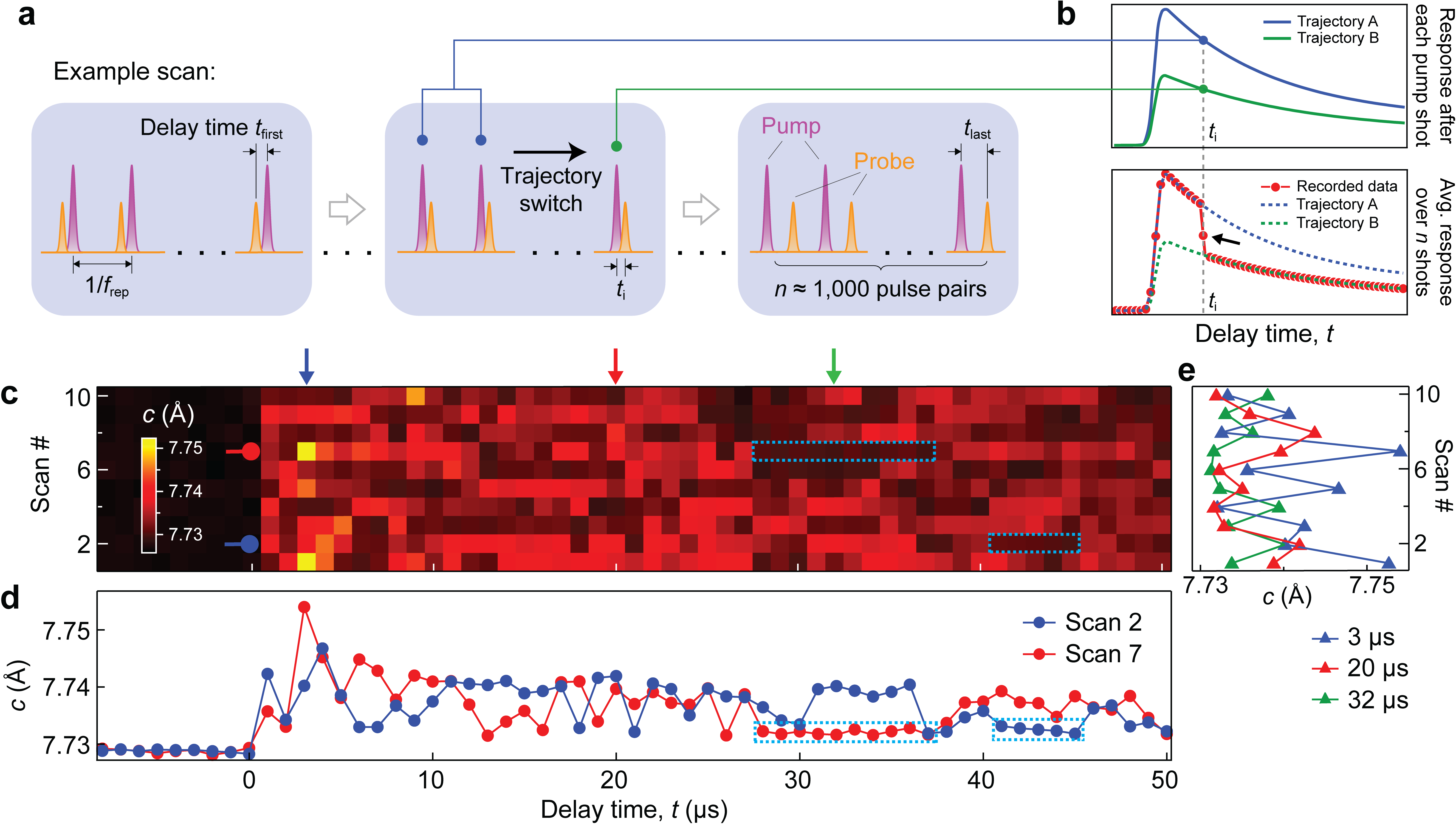}
    \caption{\textbf{Stochastic photoinduced lattice response in LLTO.} \textbf{a},~Schematic of the stroboscopic pump-probe measurement using a pump laser repetition rate of $f_\text{rep} = 1$~kHz. At each pump-probe delay time $t_{i = 1,\dots,59}$, system responses over $n \approx 1,000$ shots are accumulated. In the measurement, the pump-probe delays are changed electronically in a sequential manner, as indicated by the gray arrows. A scan consists of a complete cycle of delays $t_1,\dots,t_{59}$, and it is repeated multiple times with a negligible time overhead in between successive scans; here, only one particular scan is drawn. \textbf{b},~\textit{Upper panel:}~Schematic of stochastic lattice expansion trajectories induced by selected pump pulses, where the color-coded dots indicate the data points probed by the X-ray pulse at a particular time delay $t_i$ in a scan. Only two out of many possible trajectories are shown as examples. \textit{Lower panel:}~Schematic of recorded data for a particular scan, where a discontinuity (highlighted by the black arrow) occurs due to a stochastic change of the photoinduced lattice trajectory during data collection at delay time $t_i$. Only one such discontinuity is shown as an example, but many such discontinuities can occur during the course of one scan. \textbf{c},~Measured time evolution of the $c$ lattice parameter of a particular LLTO crystalline grain extracted from the (004) Bragg peak, shown for 10 consecutive scans under a 349-nm pump laser with a 67~mJ/cm$^2$ incident fluence at room temperature. \textbf{d},~Two representative time traces in scan~2 and scan~7 taken from panel~\textbf{c}, showing nearly identical values before $t=0$ but exhibiting dichotomous light-induced responses due to the stochastic nature of nonequilibrium lattice dynamics. Dashed boxes in \textbf{c},\textbf{d} highlight examples of a streak of similar $c$ values before or after a discontinuous shift in value. \textbf{e},~Three vertical line cuts selected from panel~\textbf{c} (see color-coded arrows), demonstrating the variation of the $c$-axis parameter across different scans at the same pump-probe delay times.}
\label{fig:2}
\end{figure*}

\section{T\lowercase{ime-resolved} X-\lowercase{ray micro-diffraction}}

A schematic of the setup is shown in Fig.~\ref{fig:1}b (see Methods for a more detailed description). As industrial-scale solid-state ionic conductors are often synthesized as sintered, pressed powder pellets \cite{Rahaman2017}, the LLTO sample under investigation was prepared in a similar polycrystalline form, whose typical grain size can be up to a few micrometers (Fig.~\ref{fig:1}c). The high momentum-resolution of the setup as well as the comparable X-ray beam spot size and LLTO grain size make it possible to observe individual Bragg peaks instead of Debye-Scherrer rings (Fig.~\ref{fig:1}d), enabling us to focus on the stochastic lattice dynamics of a single grain without an averaging effect within the powder ensemble.

To understand how the experiment is sensitive to shot-to-shot variation of the lattice response despite a conventional stroboscopic pump-probe scheme, it is worth revisiting the measurement protocol, summarized in Fig.~\ref{fig:2}a. The repetition rate of the ultraviolet pump laser was 1~kHz, and the X-ray diffraction peak was measured at each pump-probe delay time $t_{i=1,\dots,59}$ sequentially, where diffraction intensities of $n$ pump-probe pulse pairs were averaged for each $t_i$ during the delay time scan. Here, the dwell time at each delay was 1~s, leading to $n\approx 1,000$.  Upon the completion of one full pump-probe delay scan, we repeated the procedure for a total of $N$ scans ($N=10$) with nearly zero waiting time in between successive scans. In general, the sample response after every single pump laser pulse can follow different trajectories due to fluctuations (Fig.~\ref{fig:2}b, upper panel). Even though we only captured a single point out of the entire photoinduced evolution at each delay time $t_i$ (solid dots in the upper panel of Fig.~\ref{fig:2}b), as illustrated in the lower panel of Fig.~\ref{fig:2}b, we can still detect the stochastic variation between certain shots based on abrupt discontinuities in the recorded time trace (highlighted by the black arrow), especially if such discontinuities occur long after the region of pump-probe temporal overlap and if they are much larger than the noise level of the measurements (see ref.~\cite{SM} for noise estimates). 

\begin{figure*}[htb!]
    \includegraphics[width=0.70\textwidth]{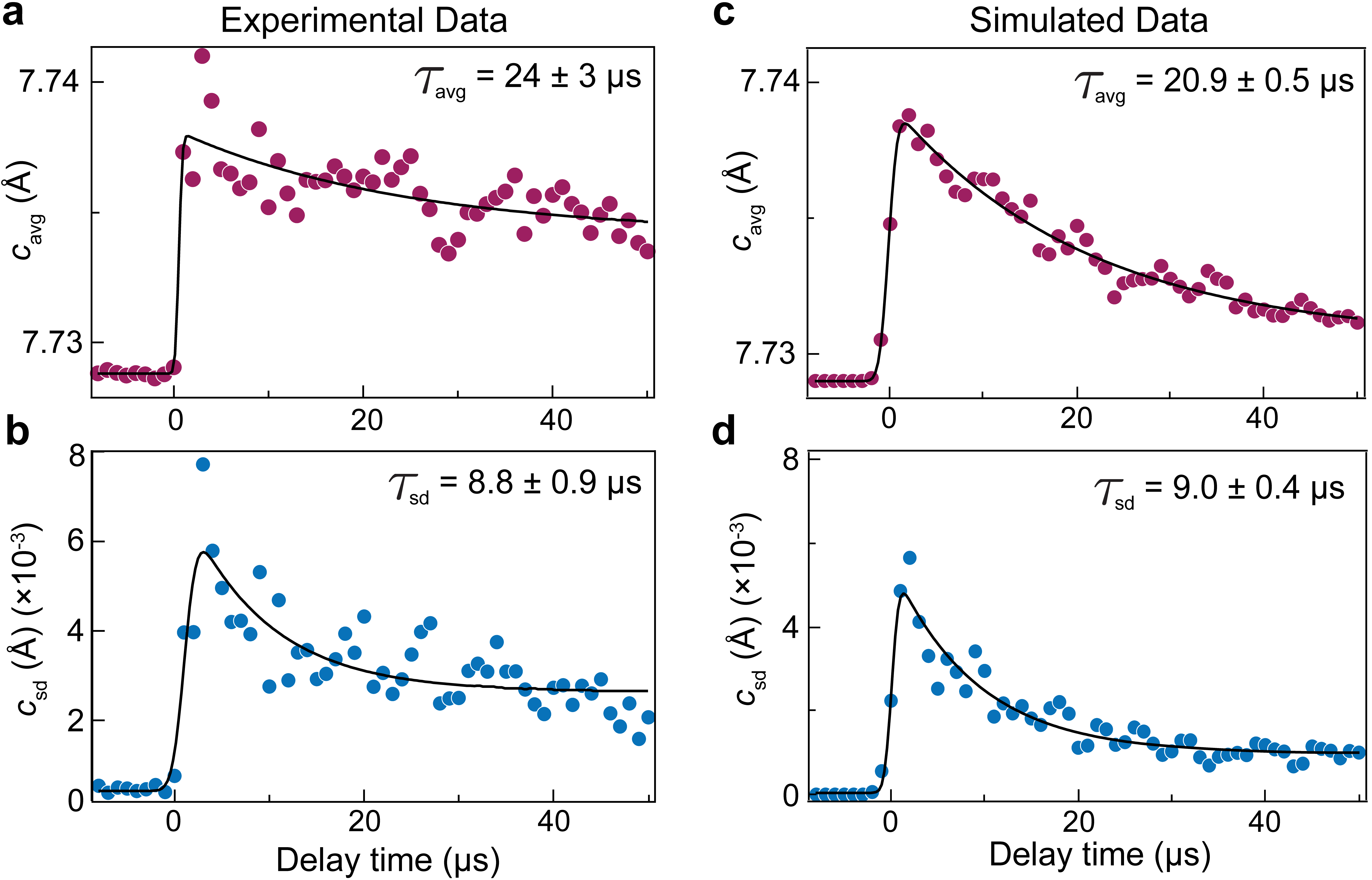}
    \caption{\textbf{Average lattice response and its fluctuations.} \textbf{a},~Time evolution of the $c$ lattice parameter averaged over the 10~scans in Fig.~\ref{fig:2}c. \textbf{b},~Time evolution of the standard deviation $c_\text{sd}$ derived from the 10 scans in Fig.~\ref{fig:2}c. In \textbf{a},\textbf{b}, the solid curves are fits to a phenomenological model in Eq.~\eqref{eq:fit}. The characteristic relaxation time $\tau_\text{sd}$ is markedly shorter than $\tau_\text{avg}$. \textbf{c},\textbf{d},~Simulated data corresponding to the measured results in panels~\textbf{a},\textbf{b}, reproducing the difference between $\tau_\text{avg}$ and $\tau_\text{sd}$. See the main text and ref.~\cite{SM} for simulation details.}
    \label{fig:3}
\end{figure*}

These discontinuities are clearly observed in our measurements. Figure~\ref{fig:2}c shows the $c$-axis lattice parameter as a function of pump-probe delay time for ten scans, two of which are shown in Fig.~\ref{fig:2}d (see ref.~\cite{SM} for the extraction of $c$ from diffraction images). Upon photoexcitation, on average, the lattice exhibits a sudden $c$-axis expansion of more than 0.1\% followed by a slow recovery over tens of microseconds (Fig.~\ref{fig:3}a). For individual scans, however, the $c$ value experiences seemingly random discontinuities in the time trace after pump pulse arrival. By contrast, the variation of $c$ before photoexcitation is much smaller (e.g., 25-times smaller in scan~7 in Fig.~\ref{fig:2}d; see ref.~\cite{SM} for noise estimates). This dichotomy of the noise level before and after $t=0$ excludes extrinsic measurement uncertainties due to factors such as instability of the X-ray beam, which is expected to yield a similar noise level at all time delays. These discontinuities hence suggest that indeed the sample response after each pump shot varies significantly, an observation further substantiated by the large contrast of the transient $c$-axis parameters at a fixed time delay across different scans (Fig.~\ref{fig:2}e).

Upon closer scrutiny, two features of the stochastic response reflected in these discontinuities stand out in Fig.~\ref{fig:2}c--e. First, the relative magnitude of a shift in $c$ is larger right after $t=0$ compared to later pump-probe delays. This feature is further echoed by the larger scan-to-scan variation at $t=3~\upmu$s (blue triangles) compared to $t=20~\upmu$s and 32~$\upmu$s (red and green triangles) in Fig.~\ref{fig:2}e. Second, the discontinuities in the time traces are often preceded or succeeded by a streak of similar values, as highlighted in dashed boxes in Fig.~\ref{fig:2}c,d. The presence of these streaks are unexpected if the stochastic lattice dynamics following each pump pulse are independent from one another. The streaks hence hint at some correlated photoinduced dynamics in LLTO. We will quantify both features of the stochastic response in the statistical analyses below.

\begin{figure*}[tb!]
    \includegraphics[width=0.83\textwidth]{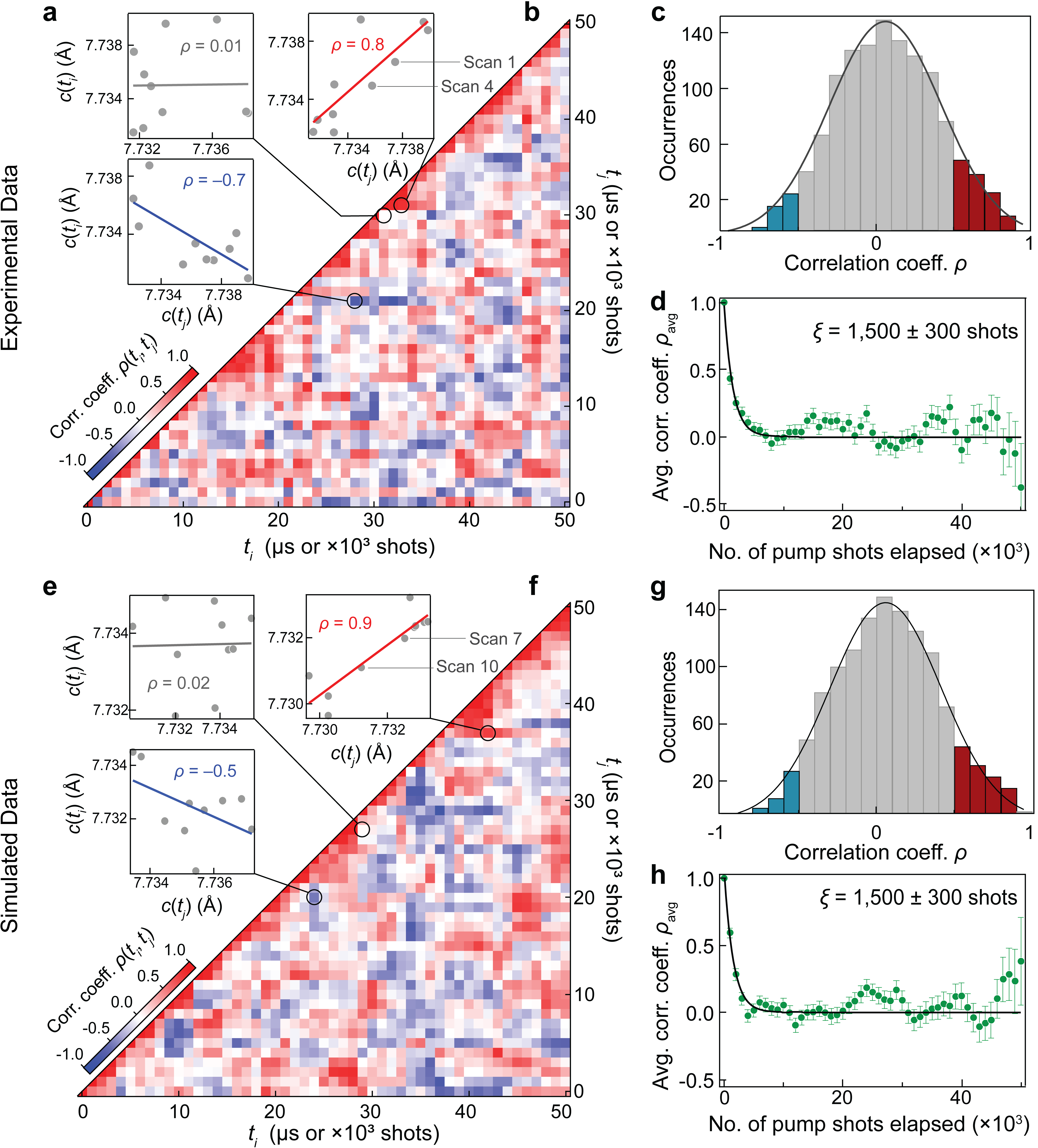}
    \caption{\textbf{Hidden correlations between the shot-to-shot responses.}  \textbf{a},\textbf{b},~Pearson correlation $\rho(t_i,t_j)$ of the measured lattice parameter $c$ between two delay times $t_i, t_j \in [0, 50]~\upmu$s, which are alternatively expressed as $10^3$~shots because 1~$\upmu$s delay step in our sequential data acquisition scheme corresponds to 1,000 pump laser shots incident on the sample. Each pixel in the correlation triangle in \textbf{b} is derived from a scatter plot between $c(t_i)$ and $c(t_j)$. Three examples of positive, nearly zero, and negative Pearson correlation coefficients are shown in \textbf{a}, where each gray dot represents the data point from one scan out of the 10 scans in Fig.~\ref{fig:2}c. The values in the correlation triangle are nearly random except for a notable positive correlation along the 45$^\circ$ diagonal where $t_i\approx t_j$. \textbf{c},~Histogram of all non-diagonal Pearson correlation coefficients in \textbf{b}, where the solid curve is a Gaussian fit. The majority of the correlation coefficients is randomly distributed around zero, but there is a slight bias towards a positive value (highlighted by red vs. blue bars outside of $|\rho| > 0.5$) due to the large, positive correlations when $t_i \approx t_j$. \textbf{d},~Averaged Pearson correlation coefficient $\rho_\text{avg}$ as a function of $|t_i - t_j|$, the latter of which is expressed as the number of pump shots elapsed. Error bars denote the standard error of $\rho_\text{avg}$ and they become larger when there are fewer number of pixels in \textbf{b} to average for large $|t_i - t_j|$. Solid curve is a fit to an exponential decay, yielding a correlation length of $\xi = 1,500\pm300$ shots. \textbf{e}--\textbf{h},~Simulated data corresponding to the measured results in panels~\textbf{a}--\textbf{d}, reproducing all key features. See the main text and ref.~\cite{SM} for simulation details.}
    \label{fig:4}
\end{figure*}

\section{S\lowercase{tatistical analysis of lattice dynamics}}

We first examine the amplitude of the scan-to-scan $c$ value variation that appears to decay as a function of delay time. Figure~\ref{fig:3}b confirms this observation, where the standard deviation $c_\text{sd}$ computed across scans exhibits a remarkably similar temporal evolution as that of the scan-averaged dynamics ($c_\text{avg}$) in Fig.~\ref{fig:3}a. Both $c_\text{avg}$ and $c_\text{sd}$ can be well fitted to a phenomenological model (black curves in Fig.~\ref{fig:3}a,b), which captures essential features of a generic photoinduced response that approximately follows first-order kinetics during long-term recovery (with time constants $\tau_{\text{avg}}$ and $\tau_{\text{sd}}$)\cite{Moore2016,Zong2021,Demsar2012},
\begin{align}
    f(t) = f_\text{equil} + \bigg[\frac{1}{2}\bigg(1+\text{erf}\bigg(\frac{2\sqrt{\ln2}(t-t_0)}{w}\bigg)\bigg)
    \notag\\
    \cdot \big(I_\infty + I_0e^{-(t-t_0)/\tau}\big)\bigg]
\label{eq:fit},
\end{align}
where $f(t)$ is the observable of interest that depends on the pump-probe delay time $t$ obtained from the electronic delay signal, and $f_\text{equil}$ is its equilibrium value prior to photoexcitation. $w$ characterizes the initial system response time, $t_0$ determines the pump pulse arrival time, $I_0$ denotes the change right after photoexcitation while $I_\infty$ denotes the value after the system partially relaxes, a process characterized by a time constant $\tau$.

On the phenomenological level, the sudden increase of $c_\text{sd}(t)$ after $t=0$ is indicative of the variation of the extent of the photoinduced lattice expansion, captured by $I_0$ in Eq.~\eqref{eq:fit}. The physical origin of such variations in $I_0$ can be the randomness of the strain environment \cite{Rubio2021} following the mechanical relaxation of neighboring grains from the previous laser shot, the changing absorbed fluence due to varying light attenuation through a grainy medium where different grain orientations in the vicinity lead to different degrees of scattering, or a combination of such factors. However, a careful comparison between the dynamics of $c_\text{avg}(t)$ and $c_\text{sd}(t)$ in Fig.~\ref{fig:3}a,b indicates that $I_0$ cannot be the only stochastic element that varies from shot to shot. Specifically, the relaxation time $\tau_\text{avg}$ is more than twice of $\tau_\text{sd}$. This difference in $\tau$ means that the $c$-axis parameter becomes more consistent in value at larger delay times, suggesting a correlation between $I_0$ and $\tau$ for individual pump shot-induced response. Indeed, when we simulate a large number of lattice responses by drawing $I_0$ randomly from a Gaussian distribution, we can reproduce the $\tau_\text{avg} > \tau_\text{sd}$ relation (Fig.~\ref{fig:3}c,d) if we impose a negative correlation between $I_0$ and $\tau$, where the exact functional form of the negative correlation is not critical (see ref.~\cite{SM} for more details of the simulation).

Next, we address the origin of the streaks in Fig.~\ref{fig:2}c,d, which suggest that variations in the lattice dynamics after each pump laser shot are not independent. To obtain a quantitative measure of the hidden dynamical correlation, we examine how the $c$-axis parameter within a scan at one particular delay time $t_i$ is correlated with its value at another delay time $t_j$. Even though we use delay times $t_{i,j}$ as a convenient notation, the correlation we are interested in is defined in between two lattice responses following two pump shots that arrive at different lab times; under our experimental condition, a delay difference $|t_i - t_j|$ of 1~$\upmu$s corresponds to approximately 1,000 pump shots elapsed (see measurement scheme in Fig.~\ref{fig:2}a). To the lowest order, we assume that a linear correlation can capture the relation, if any, between $c(t_i)$ and $c(t_j)$ across different pump shots. This assumption is expected to hold because a larger value of $c(t_i)$ at an early time delay indicates a larger initial lattice expansion, which in turn leads to a larger value of $c(t_j)$ during its microsecond relaxation period, where no coherent oscillatory dynamics were observed. 

Based on the values of $c(t_i)$ and $c(t_j)$ in Fig.~\ref{fig:2}c, we computed the Pearson correlation coefficients \cite{Boslaugh2012} $\rho(t_i,t_j)$ as a simple measure of their linear correlation: $\rho(t_i,t_j) = \text{cov}[c(t_i)c(t_j)]/(\sigma[c(t_i)]\sigma[c(t_j)])$, where the covariance (cov) and the standard deviation ($\sigma$) are computed across 10~scans. The symmetric correlation matrix is shown for its lower-half in Fig.~\ref{fig:4}b for all non-negative pump-probe time delays, where each pixel corresponds to a linear correlation coefficient derived from a scatter plot between $c(t_i)$ and $c(t_j)$ (Fig.~\ref{fig:4}a). Besides the $\rho(t_i,t_{j=i}) \equiv 1$ entries along the 45$^\circ$ diagonal, the correlation matrix in Fig.~\ref{fig:4}b appears to be populated with mostly random values around $\rho = 0$. This randomness is reflected in the histogram of all non-diagonal entries in the correlation matrix (Fig.~\ref{fig:4}c), which can be fitted to a Gaussian distribution where we notice a small offset of the histogram towards positive $\rho$, which is best seen from the unbalanced tails of the histogram highlighted by red and blue bars. The excess positive values of $\rho$ stem from the red features in the correlation matrix in the neighborhood of the 45$^\circ$ diagonal, indicating a high degree of correlation of the transient $c$-axis parameter if $t_i \approx t_j$; namely, pump-induced lattice trajectories are not independent if there are relatively few photoexcitation events elapsed between the corresponding pump pulses. To quantify the correlation length $\xi$ in terms of the elapsed pump shots during the measurement, we inspect how fast the Pearson correlation coefficient decays from 1 as one moves away from the 45$^\circ$ diagonal in the correlation matrix. In practice, we compute an averaged correlation coefficient $\rho_\text{avg}(\Delta t)$ by going through all $\rho(t_i,t_j)$ that satisfies $\Delta t = |t_i-t_j|$, shown in Fig.~\ref{fig:4}d. The resulting curve $\rho_\text{avg}$ shows a clear exponentially decaying trend towards zero, yielding a correlation length of $\xi = 1,500\pm 300$ shots, which are comparable to the number of pump shots received ($n\approx 1,000$~shots) at one specific delay during one scan in our data acquisition scheme. This statistical analysis hence gives a quantitative measure of how fast the dynamical correlation is lost as more photoexcitation events occur under repeated pump shots. 

\section{S\lowercase{imulation and discussion}}

To understand the physical origin of the correlation in the stochastic lattice expansion and relaxation, we simulated the measurements following the exact data acquisition scheme in the experiment (Fig.~\ref{fig:2}a). Inspired by the excellent fit of Eq.~\eqref{eq:fit} to the averaged response in Fig.~\ref{fig:3}a, we employed the same phenomenological model to capture the system evolution after individual photoexcitation event. In the simulation, stochastic responses are introduced by drawing the initial lattice response [$I_0$ in Eq.~\eqref{eq:fit}] randomly from a Gaussian distribution. However, if $I_0$ is both random and independent after every pump shot, there is no dynamical correlation and the resulting $\rho(t_i,t_j)$ matrix is truly randomly populated with no enhanced coefficients when $t_i \approx t_j$ (see ref.~\cite{SM} and Fig.~S5). To model the highly-correlated dynamics between neighboring shots, we introduce the following protocol: with probability $p_0 \ll 1$, $I_0$ will be randomly selected for the next pump shot; otherwise, the photoinduced dynamics for the next shot will be identical to the dynamics following the current shot. 

Despite the simplicity of our model, it recreated the key statistical properties observed in the experiment such as the Pearson correlation matrix and its histogram distribution (Fig.~\ref{fig:4}f,g), which look nearly indistinguishable from the experimental data (Fig.~\ref{fig:4}b,c). In individual time traces, important characteristics such as the discontinuities in between a stretch of continuous streaks are also clearly discernible (Fig.~S7). Similar to the experimental correlation matrix in Fig.~\ref{fig:4}d, a high and positive correlation value is observed near the 45$^\circ$ diagonal where $t_i \approx t_j$, which decays to zero for large $|t_i - t_j|$ (Fig.~\ref{fig:4}h). In the simulation, the value of $p_0$ was adjusted to yield the same correlation length of $\xi = 1,500\pm 300$ shots, leading to $p_0 = (0.09\pm0.02)\%$, where the error bar of $p_0$ is computed to correspond to the lower and upper bounds of the correlation length $\xi$.

Physically, $p_0$ represents the probability of a micro-grain of LLTO to randomly change its nonequilibrium lattice expansion and relaxation trajectories after the next pump shot. From an energetics perspective, the value of $p_0 = \exp{[-E_0/(k_BT)]}$ implies an energy barrier of $E_0 = 0.38\pm0.01$~eV, where $k_B$ is the Boltzmann constant and $T \approx 620$~K is the lattice temperature right after photoexcitation (see ref.~\cite{SM} for an estimate of the photoinduced lattice temperature rise). In our experiments, $E_0$ is the activation energy for the micro-grain to change its photoinduced lattice dynamics in the powder ensemble, where local strain, grain orientation, and non-uniform thermal gradients can all contribute to this barrier. To understand what can cause such a change of the micro-grain environment, we note that the value of $E_0$ is close to both theoretical and experimental values of the energy barrier for lithium ion migration in LLTO ($E_b \approx 0.3$--0.5~eV) \cite{Geng2009,Zhang2020,Woodahl2023}. Hence, one plausible scenario is that lithium ion displacement after each pump pulse \cite{Defferriere2022} may result in a slight variation in the metastable lattice structure, where the hopping lithium ion follows a different photoinduced trajectory due to the modified strain environment surrounding the crystalline grain of interest.

The photo-assisted lithium ion diffusion also offers an explanation to the negative correlation between the initial photoinduced lattice expansion [$I_0$ in Eq.~\eqref{eq:fit}] and the corresponding relaxation time $\tau$. Such a relation typically shows up in optical pump-probe studies of semiconductors \cite{Shkrob1998} and superconductors \cite{Gedik2004}, where in certain regimes, bimolecular recombination of electron-hole pairs or electron-electron pairs leads to a faster decay with a larger number of initially excited free carriers. However, such mechanisms are not expected to apply to the present measurements, where relaxation over tens of microseconds is dictated by heat diffusion, strain relaxation, and macroscopic grain motion \cite{Stoica2024,Cherukara2017,Chen2005}. In the context of photoinduced structural change --- especially through a nonequilibrium transition --- a positive correlation between $I_0$ and $\tau$ is typically observed \cite{Zong2019,Huber2014,Beaud2014}, contrary to our observation of LLTO. The positive $I_0$-$\tau$ correlation in those other experiments \cite{Zong2019,Huber2014,Beaud2014} is indicative of a soft mode as the system enters a flat potential energy landscape near the critical point. By contrast, in LLTO, lithium ion migration during the course of a photoinduced lattice evolution allows the lattice to enter a transient structure that is energetically more favorable than the one without any lithium ion movement. This process hence finds a local minimum by going away from the flat region in the potential energy surface, leading to a stiffer instead of softer lattice and hence accounting for the faster lattice relaxation.

Our results introduce a general statistical framework to extract stochastic fluctuations in photoinduced dynamics that can be applied to a variety of pump-probe experiments, provided that the noise in the data is dominated by intrinsic sample dynamics instead of extrinsic instrumental uncertainties. In the context of X-ray sciences, the nonequilibrium noise correlation spectroscopy demonstrated in this work presents a streamlined, complementary approach to photon correlation spectroscopy in the study of structural correlations at an ultrashort timescale while avoiding the technical complexity of implementing split X-ray pulses in a free-electron laser \cite{Osaka2016,Roseker2018,Zhu2017}. Our findings hold great potential in the active pursuit of modeling and designing spatially heterogeneous or temporally fluctuating systems that underpin most materials of both fundamental and technological interest, such as unconventional superconductors \cite{Torchinsky2013,Arpaia2019}, self-assembled nanostructures \cite{Ekiz2016}, and \textit{in operando} catalysts \cite{Johanek2004}. In intrinsically nonequilibrium context such as biomolecular processes, our approach may help identify new structural dynamics and energy flow that are critical in understanding and orchestrating the reaction pathway \cite{Bhowmick2023}. This work hence provides a powerful tool to uncover hidden correlations in otherwise random dynamics that occur at the intrinsic timescale and lengthscale of electrons, ions, and molecules.

\section{Methods}
\noindent\textbf{Time-resolved X-ray micro-diffraction}: Experiments were performed at the 7ID-C beamline at the Advanced Photon Source (APS) in Argonne National Laboratory, and the setup configurations were detailed in refs.~\cite{Zhou2022, Walko2016}. The geometry of the experiment is depicted in Figs.~\ref{fig:1}b and S1. The samples were photoexcited above the bandgap of LLTO using a 3.55~eV (349~nm) femtosecond laser source, which was produced as the second harmonic of the output of an optical parametric amplifier (OPerA Solo, Coherent Inc.) using the output of an amplified Ti:sapphire laser (Legend, Coherent Inc.). The pump laser operates at 1~kHz repetition rate, which was locked to an integer division of the synchrotron repetition rate. The probing X-ray beam from the synchrotron had a pulse duration of 100~ps operating at 6.5~MHz repetition rate. The X-ray beam was then monochromatized to an energy of 8~keV (1.5498~\AA) and focused by a pair of Kirkpatrick-Baez mirrors to a cross-sectional spot size of $12~\upmu\text{m}\times15~\upmu$m (full-width at half maximum, FWHM), which was an order of magnitude smaller than the pump beam spot size at the sample position to ensure a near-uniform photoexcitation condition in the probed area. The powder pellet was mounted at the center of rotation of a six-circle diffractometer (Huber GmbH). The X-ray diffraction signals were collected by an area detector (Pilatus 100K, DECTRIS Ltd.) that was gated to selectively record the X-ray pulse that was paired with the excitation laser pulse, rendering the overall repetition rate of the experiment to 1~kHz. The time delay between the laser pump pulse and X-ray probe pulse was varied electronically, allowing us to access a timescale up to tens of microseconds, which is relevant for the slow relaxation process pertinent to heat dissipation through a grainy pellet \cite{Han2017} as well as macroscopic grain motion as a result of laser-induced lattice parameter change.\\

\noindent\textbf{Sample preparation and characterization}:
Li$_{0.5}$La$_{0.5}$TiO$_3$ was synthesized according to the procedures in ref.~\cite{Inaguma1994}. A stoichiometric amount of La$_2$O$_3$, Li$_2$CO$_3$, and TiO$_2$ were mixed in an agate mortar and pressed into pellets under 100~MPa of pressure. The pellets were placed on a bed of sacrificial powder and calcined at 800$^{\circ}$C for 4~hr then at 1200$^{\circ}$C for 12~hr at a ramp rate of 1$^{\circ}$C/min. After the crystal structure of LLTO was confirmed with X-ray diffraction, the resulting powder was pressed into a pellet with a diameter of 6~mm and a thickness of 0.5--1~mm under 620~MPa of pressure. The pellet was subsequently annealed at 1100$^{\circ}$C for 6~hr at a ramp rate of 2$^{\circ}$C/min over a bed of its mother powder. To characterize the grain morphology, scanning electron microscopy (SEM) was performed using the SE2 detector of a ZEISS 1550VP field emission SEM with an acceleration voltage of 10~kV at 10k$\times$ magnification. As shown in Fig.~\ref{fig:1}c, the grain size of LLTO is comparable to the X-ray beam spot size for the time-resolved X-ray micro-diffraction measurements.

\section{Additional Information}

\noindent\textbf{Acknowledgements}:
We thank fruitful discussions with X.~Xu and A.~Kogar, and we appreciate the support in sample synthesis and characterization from K.~See. A.Z. acknowledges support from the Miller Institute for Basic Research in Science. J.M. acknowledges funding by the National Science Foundation (NSF-REU EEC-1852537). M.Z. acknowledges funding by the Department of Energy (DE-SC0024123). This research used resources of the Advanced Photon Source, a U.S. Department of Energy (DOE) Office of Science user facility operated for the DOE Office of Science by Argonne National Laboratory under Contract No.~DE-AC02-06CH11357. \\

\noindent\textbf{Author contributions}:
A.Z., S.C., and M.Z. conceived the project. The time-resolved X-ray measurements were conducted by A.Z., H.L., K.H.P., B.G., D.A.W., and H.W. Beamline 7ID-C at Advanced Photon Source is operated by B.G., D.A.W., and H.W. A.Z. and J.M. analyzed the data and performed model calculation. K.H.P. synthesized and characterized LLTO, and prepared the sample for the beamline experiments. Z.W.B.I. performed the scanning electron microscopy measurements of LLTO. J.M. and A.Z. wrote the manuscript with critical inputs from K.H.P., S.C., M.Z., and all other authors. The research was supervised by S.C. and M.Z.\\

\noindent\textbf{Competing interests}:
The authors declare no competing interests.\\

\noindent\textbf{Data availability}: 
All of the data and calculations supporting the conclusions are available within the article and the Supplementary Information. Additional data are available from the corresponding authors upon reasonable request.


%

\end{document}